\documentclass[floatfix,aps,pre,showpacs]{revtex4}
\usepackage{graphicx}
\usepackage{amsmath}
\usepackage{amssymb}
\usepackage{mathrsfs}

\begin{document}

\title{Glassy Behavior and Jamming of a Random Walk Process for Sequentially
	 Satisfying a Constraint Satisfaction Formula}

\author{Haijun Zhou}

\affiliation{Key Laboratory of Frontiers in Theoretical Physics and
		Kavli Institute for Theoretical Physics China, 
		Institute of Theoretical Physics, Chinese Academy of Sciences,
		Beijing 100190,	China}

\date{\today}

\begin{abstract}
Random $K$-satisfiability ($K$-SAT) is a model system for studying typical-case
complexity of combinatorial optimization. Recent theoretical and
simulation work revealed that the solution space of a random $K$-SAT formula
has very rich structures, including the emergence of solution communities
within single solution clusters.
In this paper we investigate the influence of the solution space landscape
to a simple stochastic local search process {\tt SEQSAT},
which satisfies a  $K$-SAT formula in a sequential manner.
Before satisfying each newly added clause, {\tt SEQSAT}
walk randomly by single-spin flips in a solution cluster of the old subformula.
This search process is efficient when
the constraint density $\alpha$ of the satisfied subformula is
less than certain value $\alpha_{cm}$; however it
slows down considerably as $\alpha > \alpha_{cm}$ and
finally reaches a jammed state at $\alpha \approx \alpha_{j}$.
The glassy dynamical behavior of
{\tt SEQSAT} for $\alpha \geq \alpha_{cm}$ probably is due to
the entropic trapping of various communities in the solution cluster of the
satisfied subformula. For random $3$-SAT, the jamming transition point $\alpha_j$ is
larger than the solution space clustering transition point $\alpha_d$, and
its value can be predicted by a long-range frustration mean-field theory.
For random $K$-SAT with
$K\geq 4$, however, our simulation results indicate that $\alpha_j = \alpha_d$.
The relevance of this work for understanding the dynamic properties of
glassy systems is also discussed.
\end{abstract}


\maketitle

\section{Introduction}
\label{sec:01}

The random $K$-satisfiability ($K$-SAT) problem asks to determine whether
or not a
Boolean function with $M$ clauses (each of which applying a
constraint to $K$ randomly chosen variables from a set of $N$
Boolean variables) can be evaluated to be true. This problem has been studied
extensively by computer scientists in the context of typical-case
computation complexity of NP-complete combinatorial satisfaction and
optimization \cite{Kirkpatrick-Selman-1994}. Statistical physics has
played an important role in understanding the energy landscape of
the random $K$-SAT problem, especially the (zero-energy) solution
space structure of a satisfiable random $K$-SAT formula
\cite{Monasson-Zecchina-1996,Biroli-Monasson-Weigt-2000,Mezard-etal-2002,Mezard-etal-2005,Krzakala-etal-PNAS-2007}.
It was revealed that \cite{Krzakala-etal-PNAS-2007}, as the constraint
density $\alpha \equiv M /N$ of a random $K$-SAT formula increases
from zero, the solution space of the formula experiences a series of
phase transitions before it finally becomes unsatisfiable as $\alpha$
exceeds a threshold value $\alpha_{s}$. Two very important phase
transitions in this series are the clustering or dynamic transition at
$\alpha = \alpha_{d}$ (where the solution space splits into exponentially
many dominating Gibbs states) and the condensation transition at $\alpha = \alpha_c \geq
\alpha_d$ (where the number of dominating solution Gibbs states becomes sub-exponential).
Statistical physics has also contributed a very
efficient solver, survey-propagation (SP)
\cite{Mezard-Zecchina-2002,Braunstein-etal-2005},
for the random $K$-SAT problem. SP is able to find solutions for 
large random $3$-SAT formulas with
millions of variables even when the constraint density of the formulas
approach closely the satisfiability threshold $\alpha_s$. The magic
power of SP is not yet fully understood. These and other theoretical
and algorithmic achievements make the interface between theoretical computer
science and statistical physics a prosperous research area \cite{Hartmann-Weigt-2005,Mezard-Montanari-2009}. 

As its energy landscape and solution space structure both are very complex,
the random $K$-SAT problem may serve as an interesting model system for
studying glassy dynamics. For example, similar to
lattice glass models \cite{Biroli-Mezard-2002} and spin-facilitated kinetic
Ising models \cite{Fredrickson-Andersen-1984,Ritort-Sollich-2003},
one may apply a random external field to each variable
of a random $K$-SAT formula and investigate how the configuration of the
variables changes with time under the hard constraint of the clauses (which
are not allowed to violate during the configuration evolution process).
For a single-spin-flip dynamical process, the configuration of the system
is confined to a connected component (a single solution cluster \cite{note1})
of the solution space of the $K$-SAT formula. However, the dynamics
within such a
solution cluster is not necessarily trivial. Simulations performed on
random $3$-SAT and $4$-SAT formulas revealed that, as the constraint
density $\alpha$ becomes relatively large, the structure of a single solution
cluster of the formula is no longer homogeneous \cite{Zhou-Ma-2009}. The
solutions may aggregate into many communities, each of which containing a
group of solutions that are densely connected to each other while the
connections between two different communities are relatively sparse. In a
community-rich solution cluster, a local search process will get trapped
into a solution community for some time before it passes through a bridge
region and enters into another different community. Such an entropic
trapping effect will result in multiple time scales in the relaxation dynamics,
which are typical of glassy systems.

In this work we study the dynamics of a simple process {\tt SEQSAT} 
which performs an unbiased random walk in a solution cluster of
a random $K$-SAT subformula $F_m$.  The subformula $F_m$ contains the first
$m$ clauses ($m=0$ initially) of a long random $K$-SAT formula $F$ of
$N$ variables. As soon as {\tt SEQSAT} reaches a solution of $F_m$ that
also satisfies the $(m+1)$-th clause of formula $F$, this clause is
added to subformula $F_m$ and {\tt SEQSAT} starts walking randomly in the
solution cluster of the enlarged subformula $F_{m+1}$ again.
The waiting time needed for
{\tt SEQSAT} to find a solution that satisfies the $(m+1)$-th clause
gives a measure of the viscosity in moving in the solution cluster
of the old subformula of $m$ clauses. We find that, when the constraint
density $\alpha=m/N$ of the satisfied subformula $F_m$ is small, the waiting
time to satisfy the next clause is usually small.
However the average value of the waiting time increases considerably
as $\alpha$ exceeds a threshold value $\alpha_{cm}$; and finally when
$\alpha$ reaches a larger threshold value $\alpha_j$, the average
waiting time essentially diverges and {\tt SEQSAT} stops to satisfy the
next clause.
The dramatic slowing down of {\tt SEQSAT} at $\alpha \geq \alpha_{cm}$ is
understood in terms of the complex community structures in the explored
single solution clusters of the subformula $F_m$. 
We also calculate the mean value of the jamming transition point $\alpha_{j}$
by a long-range frustration mean-field theory
\cite{Zhou-2005a,Zhou-2005b}, and find that the theoretical
prediction is in good agreement with simulation
results in the case of random $3$-SAT. For $K=3$, $\alpha_j$ is larger than the
solution space clustering transition point $\alpha_d$, while for $K\geq 4$ it
appears that $\alpha_j$  coincides with $\alpha_d$.

Recently the idea of constructing solutions for a constraint satisfaction
formula by adding constraints one after another was explored by
Krzakala and Kurchan \cite{Krzakala-Kurchan-2007}. The present
work differs from Ref.~\cite{Krzakala-Kurchan-2007}
in that the {\tt SEQSAT} process
never allows the energy of the system to increase, while the
{\tt WALKCOL} algorithm used in Ref.~\cite{Krzakala-Kurchan-2007} for
the random $Q$-coloring problem is
able to cross energy barriers. The {\tt SEQSAT} process is also different
from the {\tt CHAINSAT} process of Alava and co-authors
\cite{Alava-etal-2008}. Although {\tt CHAINSAT} also prohibits any energy
increases, it may satisfy a clause $m$ at the price of unsatisfying
a clause $m^\prime$ that was previously satisfied. In the {\tt SEQSAT}
process, however,
a satisfied clause will remain to be satisfied.
The performance of {\tt SEQSAT} may be further improved if such a
hard constraint can be made more softer by introducing a positive temperature
parameter. In this paper we also study the performance of a biased random
walk search process.

\section{The random $K$-Satisfiability problem}
\label{sec:k-sat}

A $K$-SAT formula $F$ contains $N$ variables
($i=1, 2, \ldots, N$) and $M$ clauses
($a=1,2,\ldots, M$). Each variable $i$ has a binary state $\sigma_i = \pm 1$,
and each
clause $a$ represents a constraint which involves a subset $\partial a
=\{i_1, i_2, \ldots, i_K\}$
of the $N$ variables
whose size $| \partial a | \equiv K$.
The energy of a spin configuration
$\vec{\sigma} \equiv \{ \sigma_1, \sigma_2, \ldots, \sigma_N\}$
is defined as
\begin{equation}
E(\vec{\sigma}) = \sum\limits_{a=1}^{M} E_a \ ,
\end{equation}
with the energy of clause $a$ being
\begin{equation}
	E_a = \prod\limits_{i \in \partial a} \frac{1 - J_a^i \sigma_i}{2} \ .
	\label{eq:Ea}
\end{equation}
In Eq.~(\ref{eq:Ea}), $J_a^i$ is the recommended spin value to variable
 $i\in \partial a$ by
clause $a$. If at least one of the variables $i \in \partial a$ takes the
recommended value
$\sigma_i= J_a^i$, then $E_a =0$ and clause $a$ is said to be satisfied.
Clause $a$ is violated
and $E_a=1$ if no variables $i\in \partial a$ takes the recommended value
$J_a^i$.

In constructing a random $K$-SAT formula, the $K$ variables in the set
$\partial a$ of each clause $a$ are randomly chosen from the $N$ variables,
and for each variable $i\in \partial a$
the preferred spin value $J_a^i$ is set to be $+1$ or $-1$ with equal probability.
Given a random $K$-SAT formula one needs to find, among the total number of
$2^N$ configurations, at least one configuration $\vec{\sigma}$ that
satisfies all the clauses ($E(\vec{\sigma}) = 0$), or to prove that no such
satisfying configurations exist. The random
$K$-SAT problem is an extensively studied model in the context of typical-case
computation complexity.
When the number of variables $N$ is very large, a random $K$-SAT formula has
a high probability
to be satisfiable (respectively, unsatisfiable) if the constraint density
$\alpha = M / N$
is less (respectively, greater) than certain
threshold value $\alpha_{s}(K)$. For $K=3$, Kirkpatrick and Selman estimated
$\alpha_{s} \simeq 4.2$ through extensive numerical simulations
\cite{Kirkpatrick-Selman-1994},
while the mean-field theory of statistical physics predicts that
$\alpha_{s}(3)=4.2667$ \cite{Mezard-etal-2002,Mertens-etal-2006}.
The theory of Mezard and co-authors \cite{Mezard-etal-2002} is also
applicable to the random $K$-SAT problem with $K\geq 4$.

There are many algorithms for finding solutions for a random $K$-SAT formula.
A widely used one is the branch and bound DPLL algorithm of Davis, Putnam,
Logemann, and Loveland (one can refer to the review
article~\cite{Gomes-etal-2008} for the major developments on the
 DPLL algorithm).
Frieze and Suen \cite{Frieze-Suen-1996} (see also \cite{Cocco-Monasson-2001}
and the review paper \cite{Achlioptas-2001}) showed that for random $3$-SAT with
constraint density $\alpha<3.003$, the generalized unit clause heuristic has a high
probability of constructing a solution in a
single descent of the DPLL tree. But for $\alpha>3.003$, backtracking
is needed and this may result in an exponential increase of computation time
\cite{Cocco-Monasson-2001}.
In the case of $\alpha > 3$, 
many stochastic search algorithms are able to outperform the DPLL algorithm in 
average computation time,
examples are  {\tt WALKSAT} \cite{Selman-Kautz-Cohen-1996}
(an enhanced version of {\tt RANDOMWALKSAT} \cite{Papadimitriou-1991}), 
{\tt ASAT} \cite{Ardelius-Aurell-2006},
{\tt CHAINSAT} \cite{Alava-etal-2008},
belief-propagation \cite{Pearl-1988,Kschischang-etal-2001} and its variants
(such as reinforcement \cite{DallAsta-etal-2008}),
and SP \cite{Mezard-Zecchina-2002,Braunstein-etal-2005}.

Mean-field theory of statistical physics \cite{Krzakala-etal-PNAS-2007}
predicted that the
solution space of a large random $K$-SAT formula experiences several structural
transitions as the constraint density $\alpha$ increases.
At the clustering transition point $\alpha= \alpha_d$, the solution space of
the formula splits into an exponential number of Gibbs states of
equal importance. Later at the condensation transition point
$\alpha= \alpha_c$ the solution space becomes dominated by only a few Gibbs
states (for the special case of $K=3$, the clustering and the condensation
transition coincide). It was also found that
some variables will become frozen to the same
spin value in all the solutions of a Gibbs state
\cite{Montanari-etal-2008,Ardelius-Zdeborova-2008}, this fact has
serious consequences for stochastic local search algorithms.

\section{Random walk processes}
\label{sec:seqsat}

Several slightly different random walk processes were exploited in
Refs.~\cite{Li-Ma-Zhou-2009,Zhou-Ma-2009} to study the solution space
structure of single random $K$-SAT formulas. Here, based on the same idea of
random walking, we present a
simple stochastic solver {\tt SEQSAT} for the random $K$-SAT problem.
Given a random $K$-SAT formula $F$ with $N$ variables and $M$ clauses with
index $a=1, 2, \ldots, M$, we denote by
$F_{m}$ the subformula which contains the $N$ variables and the first $m$
clauses. The constraint
density of this subformula is $\alpha = m/N$.
We set $m=0$ at time $\tau =0$ and satisfy this clause-free subformula $F_0$
by randomly picking a configuration $\vec{\sigma}(0)$ from the whole set of
$2^N$ configurations.
Suppose the $m$-th clause is first satisfied at time $\tau(m)$, and the
configuration at this time is $\vec{\sigma}\big(\tau(m)\bigr)$.
We then add the $(m+1)$-th clause, and if
this clause is satisfied by $\vec{\sigma}\bigl(
\tau(m)\bigr)$, we set 
$\tau(m+1)=\tau(m)$ and $\vec{\sigma}\big(\tau(m+1)\bigr)=
\vec{\sigma}\bigl(\tau(m)\bigr)$; otherwise we perform an {\em unbiased} random
walk of single-spin flips starting from $\vec{\sigma}\bigl(\tau(m)\bigr)$
in the solution space of subformula
$F_{m}$ \cite{note2} until a configuration $\vec{\sigma}\bigl(\tau(m+1)\bigr)$
that also satisfies the $(m+1)$-th clause is first reached at
time $\tau(m+1)= \tau(m)+\mathcal{N}_{m+1}$,
with $\mathcal{N}_{m+1} \times N$ being the total number of spin flips used to
reach $\vec{\sigma}\big(\tau(m+1)\bigr)$ from the solution
$\vec{\sigma}\big(\tau(m)\bigr)$. The waiting time of satisfying the
$(m+1)$-th clause is identified to be $\mathcal{N}_{m+1}$.

\begin{figure}[t]
\includegraphics[width=0.5\textwidth]{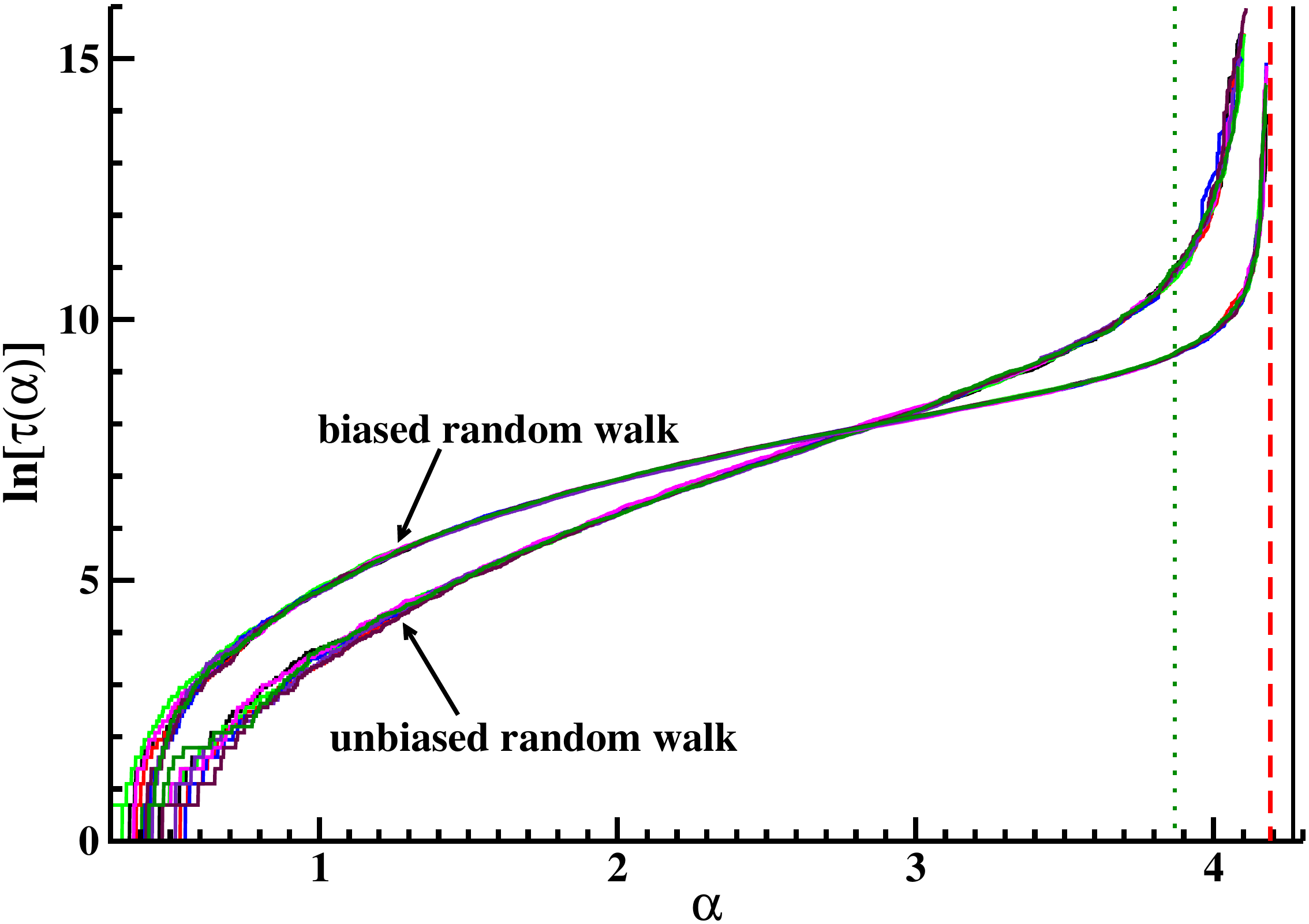}
\caption{\label{fig:3SAT}
	Logarithm of the search time $\tau(\alpha)$ by {\tt SEQSAT} to satisfy
	the first $\alpha N$ clauses of a random $3$-SAT formula of $N=10^5$
	variables. The solid lines are results obtained from
	$16$ simulation trajectories from different initial spin configurations
	(half of the trajectories are generated by an unbiased random walk rule,
	while the remaining half are by a biased random walk rule).
	The red dashed line corresponds to the jamming transition point
	$\alpha_j^\infty=4.189724$ as predicted by the
	long-range frustration mean-field theory \cite{Zhou-2005b}, the
	green dotted line marks the cluster transition point $\alpha_d$
	 \cite{Krzakala-etal-PNAS-2007}, and the right-most black solid line
	marks the satisfiability threshold $\alpha_s$ 
	\cite{Mezard-Zecchina-2002,Braunstein-etal-2005}.
	}	
\end{figure}

Starting from a random initial configuration $\vec{\sigma}(0)$, {\tt SEQSAT}
satisfies the clauses of a $K$-SAT formula $F$ in a sequential order without
overcoming any energy barriers. Simulation results of the next section
demonstrate that such a simple steepest-descent algorithm actually has
very good performance for single random $K$-SAT formulas.
There are various possibilities for further improving 
{\tt SEQSAT}. One extension is to introduce a small positive temperature
$T$ into the search process. A proposed spin flip is accepted with probability
$\exp(-\Delta E/T)$, where $\Delta E$ is the change in the configuration energy
due to this spin flip. At finite temperatures, the random walker can
overcome not only
solution space entropic barriers but also energetic barriers, therefore
its dynamics will be even richer. Another extension, which we explore in this
paper, is to use biased random walk when traveling from one solution to a
nearest-neighboring one. 

The following biased random walk jumping scheme is used.
Suppose in a configuration $\vec{\sigma}(t)$ at time $t$ there are $n$ flippable variables.
We divide these $n$ variables into two sets $A$ and $B$:
$A$ contains all the variables that are not yet flipped since being flippable
for the last time, and $B$ contains all the flippable and flipped variables.
If set $A$ is non-empty, a variable is uniformly randomly chosen from $A$ and its spin
value is flipped, otherwise, a variable in set $B$ is uniformly randomly chosen
and flipped. Once a variable becomes flippable (because all its nearest-neighbor
clauses are being satisfied by other variables), it has a larger probability to
be flipped under the biased random walk rule as compared with the previously mentioned
unbiased rule. This  gives the random walker a preference to explore new
regions of the solution space. As we show in the next section, however,
this change of local search rule does not bring in qualitative improvement
in search performance.

\section{Simulation results and interpretation}
\label{sec:simulation}

\subsection{The case of $K=3$}

\begin{figure}
\includegraphics[width=0.5\textwidth]{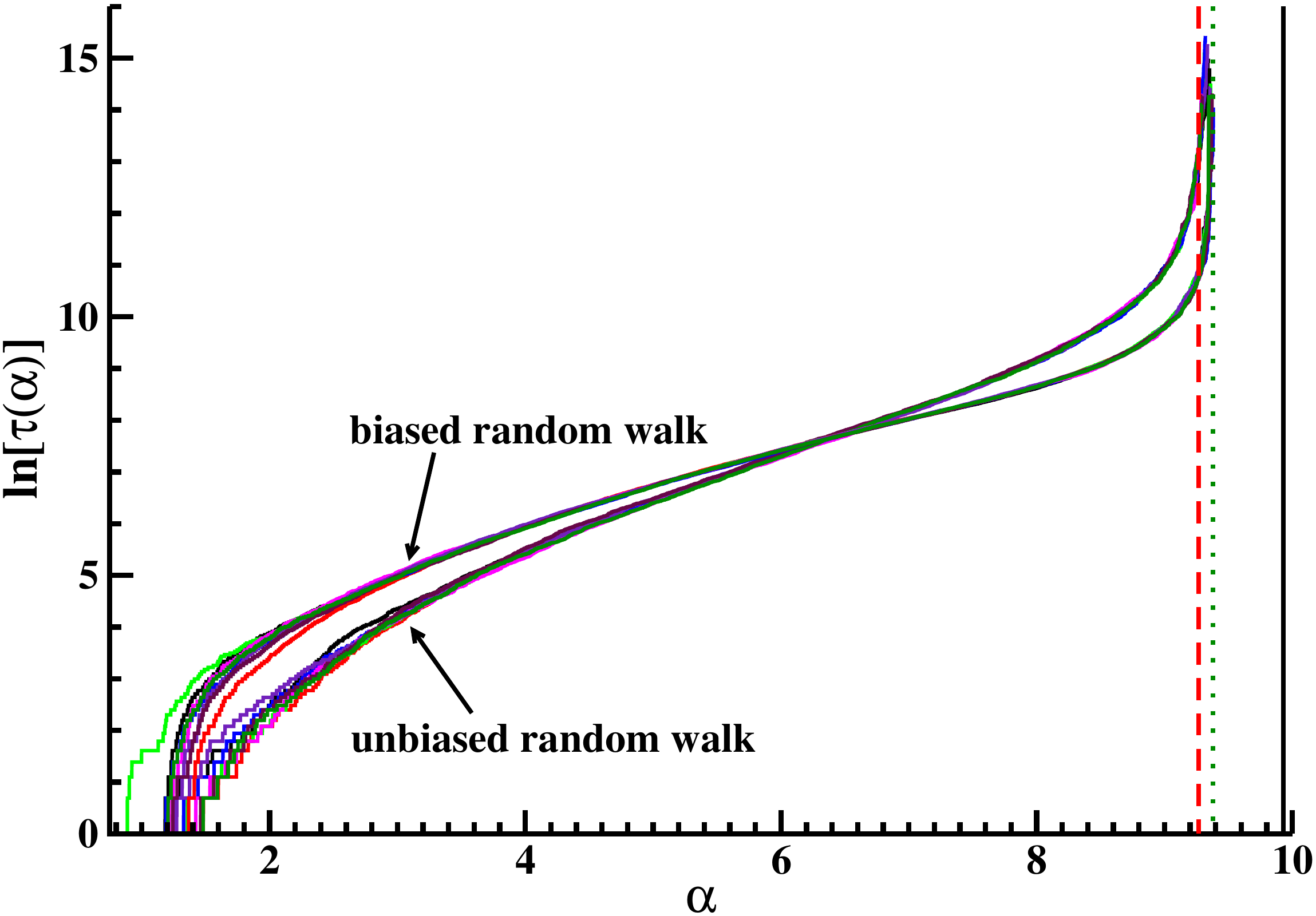}
\caption{\label{fig:4SAT}
	Same as Fig.~\ref{fig:3SAT}, but for a random $4$-SAT formula of $N=10^5$
	variables.
	}
\end{figure}

The time $\tau(\alpha)$ needed by the unbiased {\tt SEQSAT} random walk
algorithm to satisfy sequentially the first 
$\alpha N$ clauses of a random $K$-SAT formula with $N=10^5$ variables
is shown in Fig.~\ref{fig:3SAT} for $K=3$.
When the constraint density $\alpha$ of the satisfied subformula
is very low, {\tt SEQSAT} satisfy every newly added clause almost
immediately; as more clauses are added into the satisfied
subformula, {\tt SEQSAT} has to make some local adjustments of the spin
configuration to satisfy a new clause, which takes some time,
but the dynamics is still very efficient.
However as the constraint density of
the subformula is beyond $\alpha  \approx 3.8$, the unbiased {\tt SEQSAT}
slows down considerably and then essentially stops to satisfy the newly
added clause as $\alpha$ becomes even larger. For example, 
at $\alpha \approx 4.1$ it can take several weeks for the unbiased
{\tt SEQSAT} to satisfy a new clause. As a
comparison, the satisfiability threshold of the random $3$-SAT problem is
$\alpha_s \approx 4.27$
\cite{Mezard-Zecchina-2002,Mertens-etal-2006}.

\begin{figure}[t]
\includegraphics[width=0.4\textwidth]{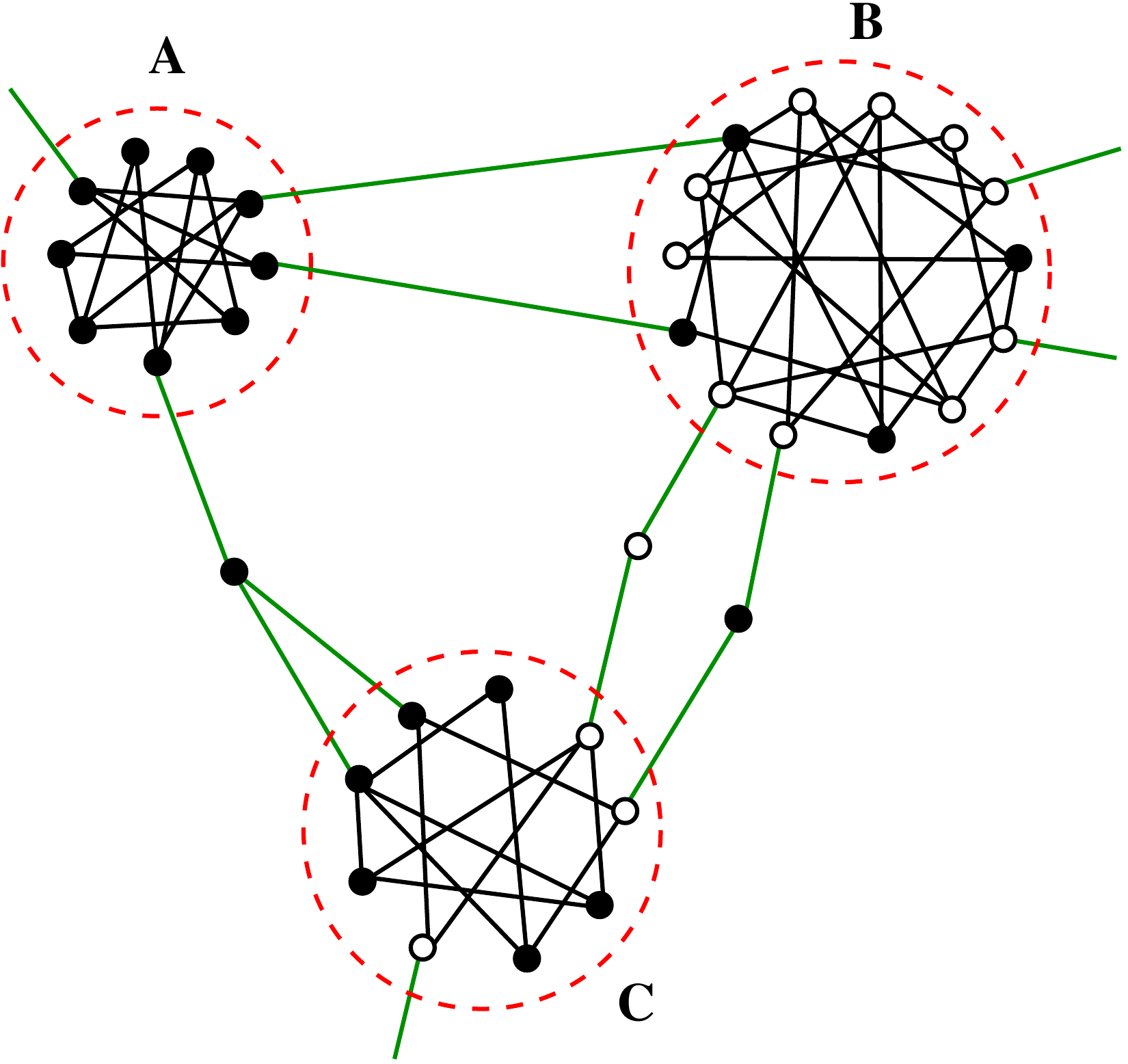}
\caption{\label{fig:community}
Schematic diagram for the connection pattern of a single solution cluster
$C_m$ of a random $K$-SAT subformula $F_m$ (containing the first $m$
clauses of a larger formula $F$). Circles represent solutions, and
edges represent a single-spin flips between two solutions of unit Hamming
distance \cite{Zhou-Ma-2009}. Those solutions that
do not satisfy the $(m+1)$-th clause of formula $F$ are denoted by filled
circles, while the remaining solutions of $C_m$ are all denoted by empty
circles. Solutions in cluster $C_m$ are grouped into many communities 
($A$, $B$, $C$, etc.) such that a solution of a community is connected to
many other solutions of the same community but are not or only very
sparsely connected to solutions of other communities. 
}
\end{figure}

We have observed the same dynamic behavior as shown in
Fig.~\ref{fig:3SAT}  when
performing the same unbiased {\tt SEQSAT} simulation on different random
$3$-SAT formulas. The search time $\tau(\alpha)$ of satisfying
the first $\alpha N$ clauses is found to scale linearly with the number
$N$ of variables when $N \geq 10^3$. Then for random $3$-SAT formulas with
different values of $N$, the curves of $\ln (\tau/N)$
as a function of $\alpha$ can be superimposed onto each other.
The diverging behavior of $\tau(\alpha)$ for
large values of constraint density $\alpha$ suggests that 
walking within a solution cluster of the subformula
$F_m$ becomes more and more viscous as $\alpha$ increases.
According to our opinion, the main reason of this viscosity increase
is the emergence of complex community structure in the solution
cluster of the subformula $F_m$ as revealed by Ref.~\cite{Zhou-Ma-2009}
(see Fig.~\ref{fig:community}). To satisfy the newly added
$(m+1)$-th clause, at least one of the attached $K$ variables of
the clause should
be flipped. However, in order to make the necessary configuration
rearrangement so that one of these $K$ variables can finally be
flipped, {\tt SEQSAT} may have to travel through many local communities
of solutions, each of them trapping the random walk process for a
period of time. Such multiple trappings will make flipping a given
variable a very difficult task.

\begin{figure}
\includegraphics[width=0.5\textwidth]{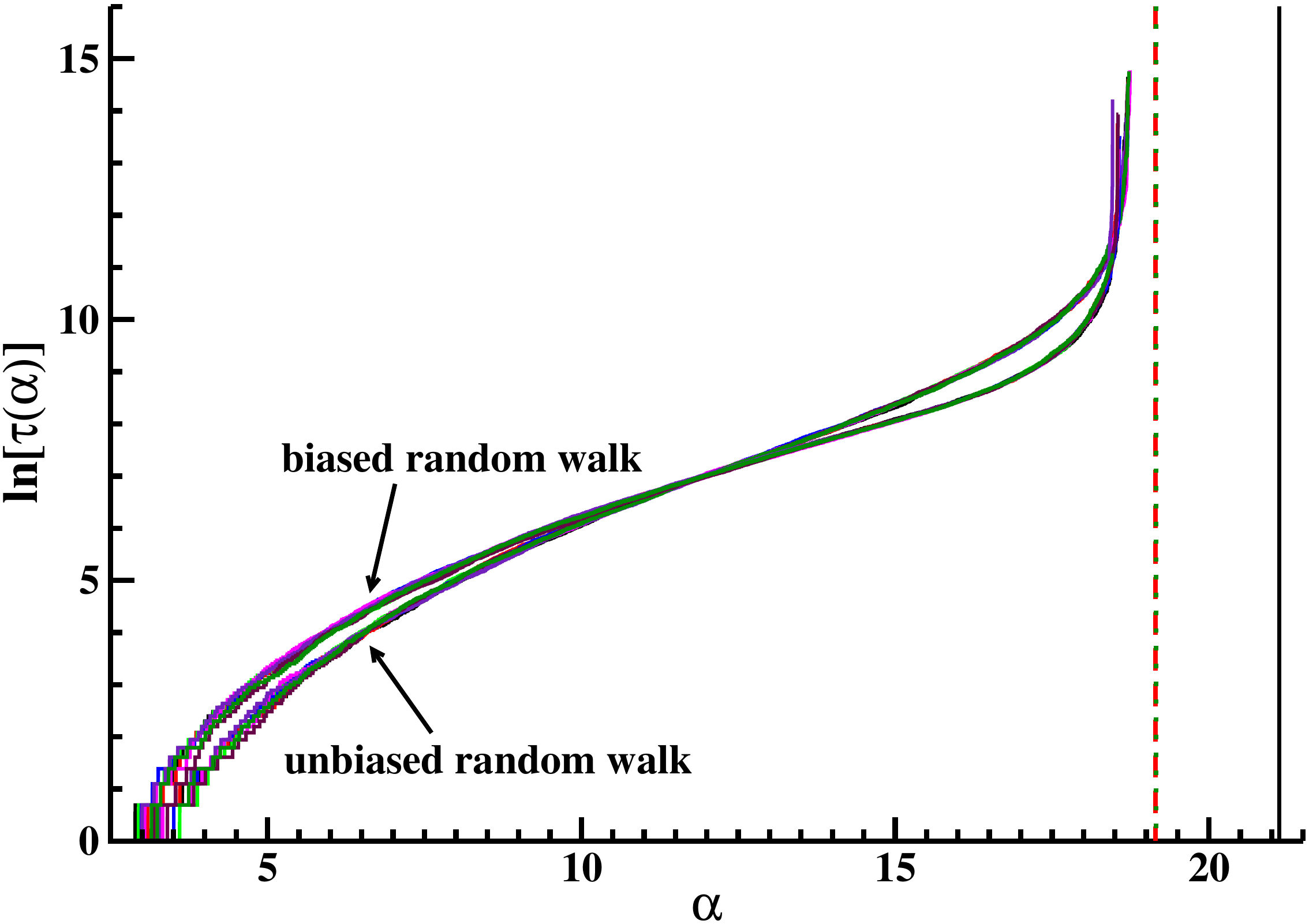}
\caption{\label{fig:5SAT}
	Same as Fig.~\ref{fig:3SAT}, but for a random $5$-SAT formula of $N=10^5$
	variables.
	}
\end{figure}

The whole solution space of a $K$-SAT formula can be represented as
a graph in which each vertex denotes a solution and each edge between
a pair of vertices means the two corresponding solutions are
related by a single-spin flip \cite{note1}.
A connected component (a solution cluster)
 of such a graph contains all the
solutions that are reachable from a reference solution by a
sequence of single-spin flips.
In the {\tt SEQSAT} process, after a solution
$\vec{\sigma}$ that satisfies the first $m$ clauses
of the random $K$-SAT formula $F$ is reached, the $(m+1)$-th clause is added
and {\tt SEQSAT} performs a random walk
of single spin flips starting from $\vec{\sigma}$ in the solution
cluster of subformula $F_m$ until a configuration
$\vec{\sigma}^\prime$ which also satisfies the newly added clause is
reached (i.e., the subformula
$F_{m+1}$ is now satisfied). Both $\vec{\sigma}$ and $\vec{\sigma}^\prime$
belong to the same solution cluster (say $C_m$) of 
subformula $F_m$. As {\tt SEQSAT} always tries to satisfy a newly added clause 
within the solution cluster of the satisfied old subformula, the
solution cluster $C_{m+1}$ of subformula $F_{m+1}$ reached by
{\tt SEQSAT} is a subset of $C_m$, i.e.,
\begin{equation}
	C_0 \supseteq C_1 \supseteq \ldots \supseteq C_m
	\supseteq C_{m+1} \supseteq \ldots \ .
\end{equation}

According to the numerical studies of
Refs.~\cite{Zhou-Ma-2009,Li-Ma-Zhou-2009},
the connection pattern of a single solution cluster $C_m$ of a random
$K$-SAT subformula $F_m$ may be quite heterogeneous. Some of the solutions
may be densely inter-connected with each other but are only very sparsely
connected to the other solutions of the same cluster. These solutions then
form a solution community (Fig.~\ref{fig:community}). Different 
communities of the same solution cluster are linked together
by inter-community edges and/or single solutions that lie at the
borders of several different communities.  In a community-rich
solution cluster $C_m$ of subformula $F_m$, some of the communities
(e.g., community $A$ in Fig.~\ref{fig:community}) may only contain solutions
that do not satisfy the $(m+1)$-th clause of formula $F$. If {\tt SEQSAT}
starts unfortunately from a solution of such a community $A$, it
then has to waste some time wondering along the internal edges of
community $A$ until it finally jumps onto an edge that leads to
another different community $B$. (Even if the community $A$ contains some
solutions that satisfy the $(m+1)$-th clause, if the fraction of such
solutions in community $A$ is very small, {\tt SEQSAT} still will
take time to reach them.)

\begin{figure}[t]
\includegraphics[width=0.5\textwidth]{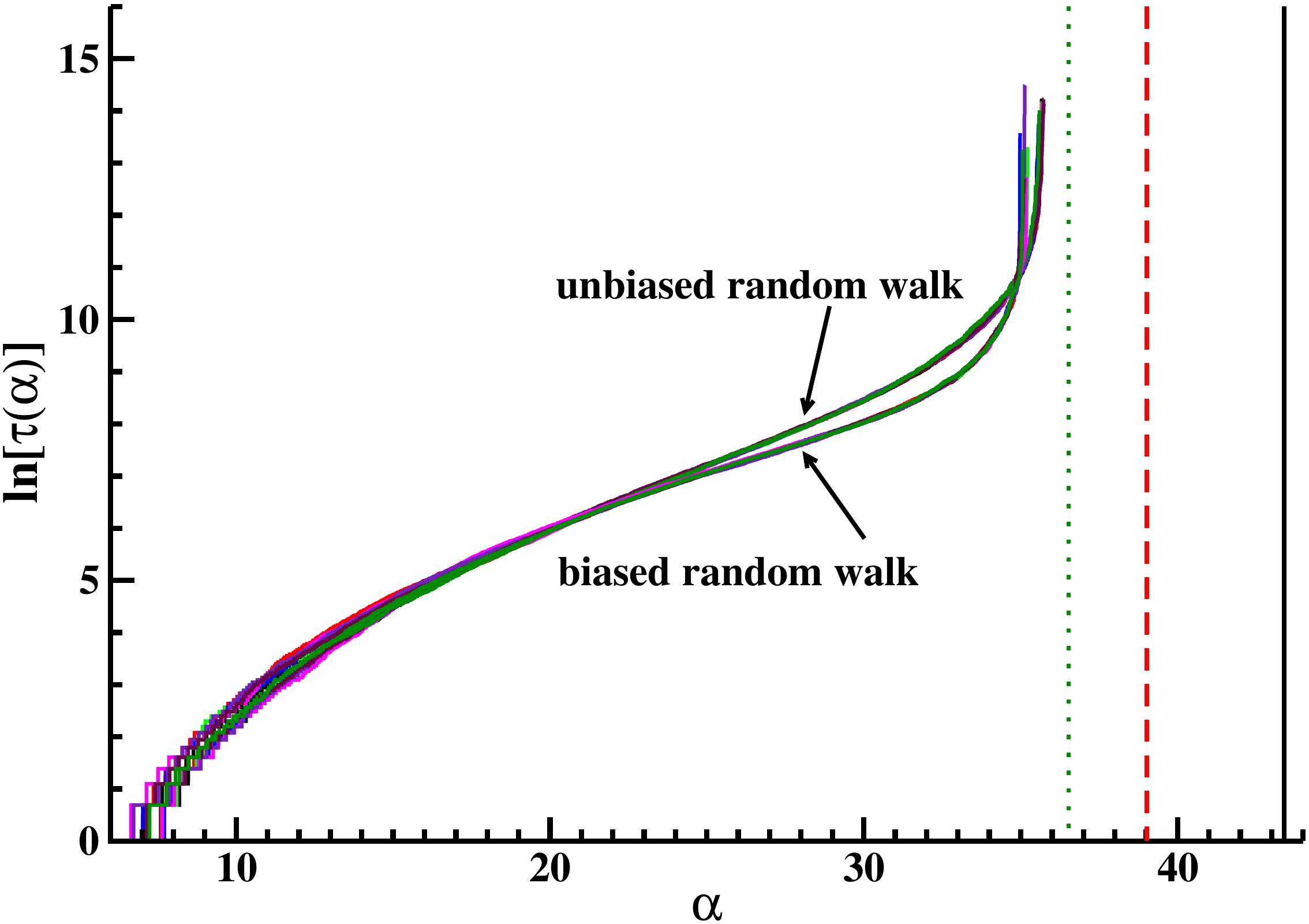}
\caption{\label{fig:6SAT}
	Same as Fig.~\ref{fig:3SAT}, but for a random $6$-SAT of $N=10^5$
	variables.
	}
\end{figure}

If the above-mentioned entropy trapping effect of solution communities
is the major reason for the slowing down of {\tt SEQSAT}, then 
the constraint density value $\alpha$ at which $\tau(\alpha)$ begins to 
increases rapidly should be close to the value of $\alpha$ at
which community structure begins to appear in a single solution cluster
of a random $K$-SAT formula.  When running {\tt SEQSAT} on random
$K$-SAT formulas, we can define an empirical threshold value
$\alpha_{cm}$ as the
{\em first} constraint density at which {\tt SEQSAT} takes 
$\mathcal{N}_{cm}$ time units (i.e., $\mathcal{N}_{cm} \times N$ single
spin flips) or more to satisfy the next clause.
Of cause $\alpha_{cm}$ is a random variable
that takes slightly different values in different trajectories
of {\tt SEQSAT} on the same formula. For the eight unbiased random walk
trajectories shown in Fig.~\ref{fig:3SAT}, we find that
$\alpha_{cm} \approx 3.37$ if we
set $\mathcal{N}_{cm}=100$ and $\alpha_{cm} \approx 3.82$ if
we set $\mathcal{N}_{cm}=1000$.  These results are
consistent with the prediction of Ref.~\cite{Zhou-2009-b}
that the solution space of the random $3$-SAT problem
is heterogeneous and community-rich at constraint
density $\alpha \geq 3.75$.  The heterogeneity of the
solution space of the random $3$-SAT problem causes the slowing down of
the unbiased {\tt SEQSAT} local search process. 

A clustering transition occurs in the
solution space of the random $3$-SAT problem at
$\alpha = \alpha_d = 3.87$, with the solution space breaks into
exponentially many solution clusters \cite{Krzakala-etal-PNAS-2007}.
As shown in Fig.~\ref{fig:3SAT}, when $\alpha> \alpha_d$
the unbiased {\tt SEQSAT} process
is still able to find solutions for a random $3$-SAT formula. The
solution space ergodicity-breaking transition therefore does not
lead to divergence of the search time of {\tt SEQSAT}. Similar
phenomena were observed in previous studies
\cite{Krzakala-Kurchan-2007,Alava-etal-2008}.

\subsection{The cases of $K\geq 4$}

The simulation results of the unbiased {\tt SEQSAT} process for random
$K$-SAT formulas are shown in Fig.~\ref{fig:4SAT} ($K=4$),
Fig.~\ref{fig:5SAT} ($K=5$) and Fig.~\ref{fig:6SAT} ($K=6$).
Similar to the results obtained for the random $3$-SAT formula,
as the constraint density $\alpha$ becomes large the {\tt SEQSAT}
process slows down exceedingly in the cases of $K\geq 4$.
Compared with Fig.~\ref{fig:3SAT} of the $K=3$ case,
the main difference for $K\geq 4$ is that
the search time of the unbiased {\tt SEQSAT} appears to
diverge at the solution space clustering transition point
$\alpha=\alpha_d$. 

This difference may be understood by recalling that at and after the clustering
transition point  $\alpha_d$, the solution
space of a random $3$-SAT formula is still dominated by a few largest clusters,
while that of
a random $K$-SAT ($K\geq 4$) formula is dominated by an exponential number of
relatively small clusters  \cite{Krzakala-etal-PNAS-2007}. The spin values
of a finite fraction
of the variables may be frozen in each of these small clusters of a
random $K$-SAT formula ($K\geq 4$), and consequently a newly added clause
has a large probability to be unsatisfied by all the solutions in the
solution cluster.

\subsection{Comparing biased and unbiased random walk search processes}

Figures~\ref{fig:3SAT}-\ref{fig:6SAT} also compare the performance of the
biased random walk search process with that of the unbiased random
walk search process. For random $3$- and $4$-SAT formulas, the biased
{\tt SEQSAT} process is more efficient than the unbiased
process, but both processes appear to diverge at the same
critical constraint density values. For random $5$- and $6$-SAT formulas,
on the other hand, the divergence point of the biased {\tt SEQSAT} process
is smaller than that of the unbiased process.

\section{Long-range frustration theory on the jamming transition}
\label{sec:jamming}

The simulation results of the preceding section demonstrate that,
the random walk searching process {\tt SEQSAT}, being prohibited from
jumping between different solution clusters of a $K$-SAT subformula
$F_m$, will eventually reach a jammed state
after the first $m_{max}= \alpha_j N$
clauses of the original random formula $F$ have been satisfied. {\tt SEQSAT}
is unable to satisfy the $(m_{max}+1)$-th clause if in the
reached solution cluster of $F_{m_{max}}$, the variables that are involved
in this clause are all frozen to the $`$wrong' spin value.
The critical constraint density $\alpha_j = m_{max}/N$ 
(the jamming transition point) may be different in
different runs of {\tt SEQSAT}, as  different runs of {\tt SEQSAT} may
reach different solution clusters of the solution space. Our simulation
results reveal that the jamming constraint densities $\alpha_j$ as obtained
for many trajectories of {\tt SEQSAT} on the same random $K$-SAT
formula and on different random $K$-SAT formulas are very close to each other.
It is anticipated that, in the thermodynamic limit of $N\rightarrow \infty$,
the {\tt SEQSAT} process has a true jamming transition at a
critical constraint density $\alpha_j^\infty$. 

The jamming transition of {\tt SEQSAT} is closely related to the freezing of
variables in a solution cluster of a random $K$-SAT formula. Recently there
were several interesting studies on the freezing transition of the solution
space of the random $K$-SAT problem
\cite{Zdeborova-Krzakala-PRE-2007,Semerjian-2008,Ardelius-Zdeborova-2008,Montanari-etal-2008}. For example, the freezing transition point for the random
$3$-SAT problem is located at $\alpha=\alpha_f= 4.254$
\cite{Ardelius-Zdeborova-2008},
while that for the random $4$-SAT problem is located at
$\alpha_f=9.88$ \cite{Montanari-etal-2008}. These threshold constraint
densities correspond to the appearance of frozen variables in the dominating
Gibbs states of the solution space. On the other hand, the final solution
clusters reached by the {\tt SEQSAT} process should not be the
dominating clusters of the solution space (otherwise, the
$(m+1)$-th clause would be satisfiable). Therefore it is natural
to expect (and is confirmed by simulation results shown in
Figs.~\ref{fig:3SAT} and \ref{fig:4SAT}) that the jamming transition
point $\alpha_j^\infty$ of {\tt SEQSAT} is less than the freezing transition
point $\alpha_f$. 

As shown in Figs.~\ref{fig:4SAT}-\ref{fig:6SAT}, the jamming transition
point $\alpha_j$ for random $K$-SAT formulas with $K\geq 4$ are very close to the
critical value $\alpha_d$. On the other hand, for random $3$-SAT formulas, $\alpha_j$
is much larger than $\alpha_d$. 
We now try to predict the value of $\alpha_j^\infty$ as a function of $K$
using the long-range frustration theory of
Refs.~\cite{Zhou-2005a,Zhou-2005b}. In a solution cluster of a random
$K$-SAT formula with $N\rightarrow \infty$ variables and constraint density
$\alpha$, some of the variables are unfrozen as their spin values are positive
in some of the solutions of this cluster and negative in the remaining
variables. We denote the fraction of
unfrozen variables in this solution cluster as $q_0$. With respective to a
pre-specified spin value $\sigma_i^*$, a unfrozen variable $i$ can be
regarded as either type-I unfrozen or type-II unfrozen \cite{Zhou-2005b}.
If flipping variable $i$ to $\sigma_i=\sigma_i^*$ leads to the
fixation of the spin values of a finite fraction
of all the other unfrozen variables, then variable $i$ is type-I unfrozen;
if setting $\sigma_i=\sigma_i^*$ only affects a small number of other
unfrozen variables, then $i$ is type-II unfrozen. We denote by
$R$ the probability that a randomly chosen unfrozen variable $i$ is
type-I unfrozen with respective to a randomly specified
spin value $\sigma_i^*$.

The spin states of the
type-I unfrozen variables of the random $K$-SAT formula are strongly
correlated.
If one fix the spin of one such variable, the final effect might be
that the spins of a large fraction of all the other
type-I unfrozen variables are also being fixed.
A type-I unfrozen variable takes
different spin values in different dominating communities, but within
each community its spin is frozen to one value.

As one adds more clauses to the random $K$-SAT formula, the solution
cluster shrinks, and then both $q_0$ and
$R$ will change. According to Ref.~\cite{Zhou-2005b} the following
set of self-consistent equations can be derived:
\begin{eqnarray}
q_0 & = & \sum\limits_{m=0}^{\infty} \bigl( P_{v}(m) \bigr)^2 \ , 
	\label{eq:q0} \\
R   & = & 1-\exp(-\lambda_{3} R)  \ . \label{eq:r}
\end{eqnarray}
In the above two equations, the function $P_{v}(m)$ is defined by
\begin{equation}
	P_{v}(m)= \sum\limits_{n=0}^{m} f(n, \lambda_2) P_{f}(m-n) 
\end{equation}
with
\begin{eqnarray}
	P_{f}(n) &=& f(2n, \lambda_1) C_{2 n}^n 2^{-2 n}
	 + \sum\limits_{s=2 n +1}^{\infty}
	f(s, \lambda_1) C_{s}^{n} 2^{1- s} \ , \nonumber \\
	f(n, \lambda) & = & e^{-\lambda} {\lambda}^n / n! \ ,
	\nonumber
\end{eqnarray}
and the three $\lambda$ parameters are expressed as
\begin{eqnarray}
	\lambda_{1} &=&
	 K \alpha \bigl( ((q_0 R+1-q_0)/2)^{K-1} - ((1-q_0)/2)^{K-1} \bigr)
	\ ,
	\nonumber	\\
	\lambda_{2} &=& (K \alpha/ 2) ((1-q_0)/2)^{K-1}  \ ,
	\nonumber	\\
	\lambda_{3} &=& (K(K-1)  \alpha q_0 (1-R) /2) ((1-q_0)/2)^{K-2}
	\ . \nonumber
\end{eqnarray}
\begin{table}[ht]
	\begin{tabular}{l|l|l|l|l}
	\hline
	\hline 
	$K$	&	$\alpha_{j}^\infty$	& $\alpha_{d}$	\cite{Krzakala-etal-PNAS-2007} & $\alpha_{s}$ \cite{Mertens-etal-2006}	& $q_0^{j}$ \\
	 \hline
	$3$ 	&	$4.1897$		& $3.87$	& $4.2667$	& $0.5270$ \\
	$4$ 	&	$9.2653$		& $9.38$	& $9.931$	& $0.3994$ \\
	$5$ 	&	$19.1480$		& $19.16$	& $21.117$ \;\;\;	& $0.3359$ \\
	$6$ 	&	$39.0269$		& $36.53$ \;\;\;	& $43.37$	& $0.2967$ \\
	$7$ 	&	$79.4245$		& 		& $87.79$	& $0.2694$ \\
	$8$ 	& 	$161.78$		& 		&		& $0.2479$ \\
	$9$ 	&	$329.704$		&		&		& $0.230$ \\
	$10$	&	$671.796$		&		&		& $0.2147$ \\
	$11$	&	$1368.01$		&		&		& $0.2015$ \\
	$12$ \;\;\;	&	$2783.8$ \;\;\;		&		&		& $0.190$ \\
	\hline
	\hline
	\end{tabular}
	\caption{\label{tab:tab1}
	$\alpha_{j}^\infty$ is the threshold value of
	jamming transition as predicted by the long-range frustration
	theory \cite{Zhou-2005b}, $\alpha_{d}$ is the clustering
	transition point reported
	by Ref.~\cite{Krzakala-etal-PNAS-2007}, $\alpha_s$ is the
	satisfiability transition point \cite{Mertens-etal-2006}, and
	$q_0^j$ is the fraction of unfrozen variables at the jamming
	transition as predicted by the long-frustration frustration
	theory.}
\end{table}

The jamming transition point $\alpha_j^\infty$ corresponds to the smallest
value of $\alpha$ at which a fixed point $q_0< 1$ of Eqs.~(\ref{eq:q0}) and
(\ref{eq:r}) first appears. Table~\ref{tab:tab1} lists the value of
$\alpha_j^\infty$ for $3 \geq K\leq 12$ and the corresponding fraction of
unfrozen variables $q_0^j$ at the jamming transition. 
As a comparison with
simulation results, we have denoted by a red dashed line the
predicted jamming transition
point $\alpha_j^\infty$ in Figs.~\ref{fig:3SAT}-\ref{fig:6SAT}.
For $K=3$, we find
$\alpha_j^\infty=4.19$, which is larger than the clustering transition point
$\alpha_d=3.87$ but is in agreement with the simulation results of
Fig.~\ref{fig:3SAT}; while for $K=4$, $\alpha_j^\infty=9.27$ is
smaller than the dynamic transition $\alpha_d=9.38$. As Fig.~\ref{fig:4SAT}
shows, for random $4$-SAT formulas
the {\tt SEQSAT} process is able to reach constraint density
values higher than $\alpha_j^\infty$. 
The long-range frustration mean-field theory 
appears to give a satisfactory prediction of the jamming transition
point of {\tt SEQSAT} for random $3$-SAT formulas but fails in the case of
$K =4$ and $K=6$. 

The solution space structure of the random $3$-SAT problem is
qualitatively different from that of the random $K$-SAT
problem with $K\geq 4$ \cite{Krzakala-etal-PNAS-2007}. Beyond
the clustering transition point $\alpha_d$, the solution space
of a large random $3$-SAT formula is dominated by only a sub-exponential
number of solution clusters, while that of a large random $K$-SAT ($K\geq 4$)
is divided into an exponential number of solution clusters of equal
statistical importance. 
The full-step replica-symmetry-breaking mean-field theory is
needed to fully describe the statistical property of the solution space
of the random $3$-SAT problem 
$\alpha>\alpha_d(3)$ \cite{Montanari-etal-2004},  but for the random
$K$-SAT problems with $K\geq 4$,
a simpler first-step replica-symmetry-breaking theory is believed to be
sufficient. 
As the solution space of a random $K$-SAT ($K\geq 4$) formula has exponentially
many communities or clusters
at the vicinity of the clustering transition $\alpha_d$,
some of the assumptions of the long-range frustration mean-field theory
may no longer be appropriate.

\begin{figure}
	\includegraphics[width=0.5\textwidth]{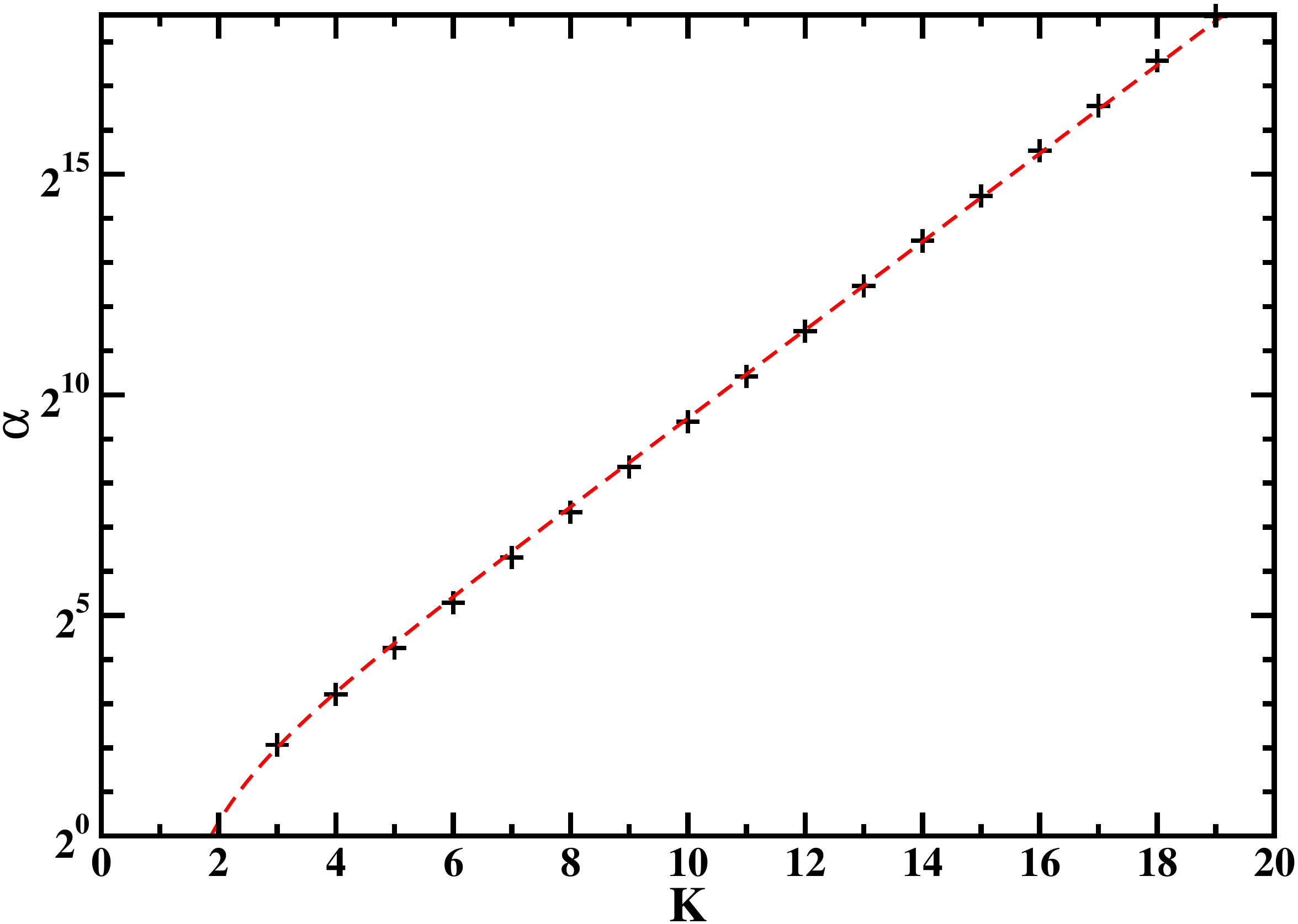}
\caption{
	\label{fig:alpha_j}
	The scaling behavior of the predicted jamming transition
	point $\alpha_j^\infty$ with $K$. The red dashed line
	is a fitting curve of the form $\alpha_j^\infty(K)= 2^K \ln 2 - c_0$,
	with the fitting parameter being $c_0=-1.5\pm 0.2$.
}
\end{figure}

Figure~\ref{fig:alpha_j} shows how the predicted jamming
transition point $\alpha_j^\infty$ scales with $K$. The data is
consistent with 
\begin{equation}
	\label{eq:scaling-alpha-j}
	\alpha_j^\infty(K)= 2^K \ln 2 + O(1) \ .
\end{equation}
Notice that the satisfiability threshold $\alpha_s(K)$ also the
same scaling behavior \cite{Mertens-etal-2006}.
The jamming value $\alpha_j^\infty(K)$
may serve as a good lower bound for the satisfiability threshold
of the random $K$-SAT problem.

\section{Conclusion and further discussions}
\label{sec:conclusion}

In this work the dynamic behavior of a simple stochastic search algorithm
{\tt SEQSAT} for the random $K$-SAT problem were investigated by computer
simulations. This simple algorithm is able to find solutions for a large
random $K$-SAT ($K\geq 3$) formula if the constraint density $\alpha$
is less than certain threshold $\alpha_j(K)$, but it
experiences a jamming transition as $\alpha$ approaches $\alpha_j(K)$ from
below.  For $K\geq 4$, we found that the jamming point
$\alpha_j(K)$ is very close to
the solution space clustering transition point $\alpha_d(K)$, but
the jamming point $\alpha_j(3) \approx 4.19$ for the special case of $K=3$
exceeds $\alpha_d(3)=3.87$ considerably. 
We argued in this work that, the dramatic slowing down of {\tt SEQSAT}
at $\alpha$ close to the jamming point
$\alpha_j(K)$ is caused by the entropic trapping effect
of various solution communities in a single solution cluster of the
random $K$-SAT formula. We also estimated the
jamming transition point $\alpha_j(K)$ using the
mean-field long-range frustration theory of
Refs.~\cite{Zhou-2005a,Zhou-2005b}, and found that
the calculated value of $\alpha_j(3)$ is in good agreement with
simulation results.

The rapid increase of the search time $\tau(\alpha)$ of the dynamic
process {\tt SEQSAT} at $\alpha$ close to the jamming point $\alpha_j$
is reminiscent of the rapid increase of viscosity of a glass-forming
liquid at low temperatures. These glassy behaviors may be governed to
a large extent by the same physical mechanisms. 
 The random $K$-SAT problem might serve as a very rich model system to
study the connection between complex energy landscapes and
glassy dynamics. In the present
paper, the dynamics of {\tt SEQSAT} is confined to a single
connected component of the zero-energy ground-state configuration space
of a $K$-SAT formula; in future studies, one may introduce external fields
and/or a finite temperature to the system to observe more complex
dynamic behaviors.

When the constraint density $\alpha$ of the random $K$-SAT formula is
slightly beyond the jamming point $\alpha_j$, the formula still
contains exponentially many solutions, but {\tt SEQSAT} is unable to
reach any one of them. On the algorithmic side, a major limitation of
{\tt SEQSAT} is that it only explores a single connected component
of the solution space. One may incorporate the energy-barrier
crossing techniques of other heuristic algorithms into
{\tt SEQSAT} to enhance its performance. We also demonstrated that
the efficiency of
searching within a solution cluster can be elevated to some extent
by using biased random walks \cite{Li-Ma-Zhou-2009,Zhou-Ma-2009},
but such a small change of local search rule does not
lead to a shift of the jamming point $\alpha_j(K)$ to larger values.

Although the mean-field long-range frustration theory \cite{Zhou-2005b}
is able to give good predictions on the jamming transition point
$\alpha_j(K)$ for $K=3$, it fails to do so for $K=4, 6$. For $K\geq 4$
the jamming transition point of {\tt SEQSAT} probably is identical to
the clustering transition point $\alpha_d$.
There is an important uncontrolled approximation in the mean-field theory,
namely that two type-I unfrozen variables have probability {\em one-half}
of being prohibited from taking simultaneously their canalizing spin values
\cite{Zhou-2005a,Zhou-2005b}. This approximation may not  be very
appropriate for the case of $K> 3$. On the hand, as shown in
Table~\ref{tab:tab1}, the predicted jamming transition
point $\alpha_j(K)$, whose value is close
to $\alpha_d(K)$, is always lower than the satisfiability threshold
$\alpha_s(K)$ and has the same scaling behavior of
$\alpha_j(K) \approx  2^{K} \ln(2)$ as $\alpha_s(K)$.

\section*{Acknowledgement}

HZ thanks Haiping Huang, Kang Li, Hui Ma, Ying Zeng, Pan Zhang, and Jie Zhou
for many helpful discussions. This work was partially supported by the
National Science Foundation of China
(Grant number 10774150) and the China 973-Program (Grant number 2007CB935903).


\begin{thebibliography}{38}
\bibitem[{not({\natexlab{a}})}]{note1}
\bibinfo{note}{The solution space of a random $K$-SAT formula can be
  represented by a graph of vertices and edges. Each solution of the formula
  corresponds to a vertex of the graph, and two vertices are connected by an
  edge in the graph if and only if the corresponding two solutions can be
  inter-changed by performing a single spin flip (i.e., they differ at only one
  variable). A solution cluster includes all the vertices (solutions) of a
  connected component of this graph. The graph may have exponentially many
  solution clusters.}
\bibitem[{not({\natexlab{b}})}]{note2}
\bibinfo{note}{Suppose the configuration of the $N$ variables is
  $\vec{\sigma}(t)=\{ \sigma_1(t), \sigma_2(t), \ldots, \sigma_N(t)\}$ at time
  $t$ and suppose that $n$ variables $\{i_1, i_2, \ldots, i_n \}$ are flippable
  in this configuration (that is, if the spin of any of the variables $i \in \{
  i_1, i_2, \ldots, i_n \}$ is fliped to $\sigma_i = - \sigma_i(t)$, the new
  configuration is still a solution of the formula). The unbiased random walk
  consists in choosing a variable in this flippable variables set with uniform
  probability and fliping its spin value.}
\end{thebibliography}

\end{document}